# Room-temperature on-chip orbital angular momentum single-photon sources

Cuo Wu,[1,2,†] Shailesh Kumar,[2,†] Yinhui Kan,[2,3] Danylo Komisar,[2] Zhiming Wang,[1] Sergey I. Bozhevolnyi,[2] and Fei Ding[2,*]

[1]Institute of Fundamental and Frontier Sciences, University of Electronic Science and Technology of China, Chengdu 610054, China

[2]Centre for Nano Optics, University of Southern Denmark, Campusvej 55, Odense M DK-5230, Denmark

[3]College of Astronautics, Nanjing University of Aeronautics and Astronautics, Nanjing 210016, China

*Corresponding author. Email: feid@mci.sdu.dk

†These authors contributed equally: Cuo Wu and Shailesh Kumar.

**Abstract:** On-chip photon sources carrying orbital angular momentum (OAM) are in demand for high-capacity optical information processing in both classical and quantum regimes. However, currently-exploited integrated OAM sources have been primarily limited to the classical regime. Herein, we demonstrate a room-temperature on-chip integrated OAM source that emits well-collimated single photons, with a single-photon purity of $g^{(2)}(0) \approx 0.22$, carrying entangled spin and orbital angular momentum states and forming two spatially-separated entangled radiation channels with different polarization properties. The OAM-encoded single photons are generated by efficiently outcoupling diverging surface plasmon polaritons excited with a deterministically positioned quantum emitter *via* Archimedean spiral gratings. Our OAM single-photon sources bridge the gap between conventional OAM manipulation and nonclassical light sources, enabling high-dimensional and large-scale photonic quantum systems for information processing.

Angular momentum that includes the spin angular momentum (SAM) (*1*) related to circular polarizations and the orbital angular momentum (OAM) associated with helical wavefronts is one of the fundamental physical properties of light. In contrast to the SAM that only takes two values of $\pm \hbar$, the OAM can carry discrete angular momenta of $\ell\hbar$ (*2-4*), where $\ell$ is an unbound integer corresponding to the topological charge and $\hbar$ is the reduced Planck constant. The possibility of generating an infinite number of states characterized by different OAMs opens up an unprecedented opportunity for high-capacity information processing in both classical and nonclassical regimes. Typically, OAM beams are generated by combining external light sources and extra phase transformers, including either bulk optical components such as computer-generated holograms (*5*) and spiral phase plates (*6*), or recently developed compact metasurfaces (*7*) and silicon resonators (*8*). Lately, integrated chip-scale microlasers have been extensively investigated for internally excited vortex beams with high purity (*9-14*).

Despite certain achievements, the insofar demonstrated on-chip OAM sources are limited to the classical regime, while compact quantum sources of OAM encoded single photons still remain largely unexplored. To realize quantum sources of photons carrying OAM, separate single photons from spontaneous parametric down-conversion (SPDC) process are combined with extra phase modulators (*15, 16*), resulting in bulky configurations and additional losses. Importantly, SPDC-based single-photon sources are inherently probabilistic, exhibiting an intrinsic trade-off between their efficiency and photon purity, a circumstance that limits their potential for being exploited in large-scale and high-dimensional photonic quantum systems. Very recently, solid-state sources for single photons with OAM have been realized using epitaxial quantum dots embedded in microring resonators cooled to a temperature of 30 K (*17*), i.e., under experimental conditions that are somewhat restrictive from the viewpoint of practical applications. Room-temperature on-chip integrated sources of single photons carrying OAM states have remained so far elusive while being in demand for high-capacity quantum information processing to meet requirements of increasing both photon states and entanglement dimensionality.

Herein, we employ deterministically positioned nano-diamonds (NDs) containing individual nitrogen-vacancy (NV) centers (ND-NVs) as single quantum emitters (QEs) to spontaneously trigger collimated single photons carrying SAM and OAM superposition states by efficiently outcoupling QE-excited diverging surface plasmon polaritons (SPPs) with an Archimedean spiral grating atop a dielectric-spacer-coated silver substrate. The considered configuration enables room-temperature generation of oppositely spinning photons carrying the conserved total angular momentum and propagating in spatially-separated entangled radiation channels.

The operation of our OAM single-photon source is schematically visualized using layer-by-layer representation in Fig. 1A. A pre-selected ND-NV with desired single QE characteristics is deterministically located in the center of an Archimedean spiral grating made of hydrogen silsesquioxane (HSQ) atop a thin silica (SiO$_2$) spacer film supported by a silver (Ag) substrate. Upon excitation with a tightly focused radially-polarized pump laser beam producing a strong longitudinal electric field component $E_z$ at the focal plane (not shown in the schematic), the QE dipole component along the $z$-axis is selectively excited, radiating primarily into cylindrically diverging (radial) SPPs supported by the composite air-SiO$_2$-Ag interface (*18*). Note that it is only the SPP in-plane components that can be efficiently outcoupled into far-field radiation. These in-plane SPP field components can be expressed in the propagation plane as follows, considering only the main terms at distances far away from the source point (*19*): $\vec{E}_{spp} = \frac{\exp(ik_{spp}r)}{\sqrt{r}}\begin{pmatrix}\cos\varphi\\\sin\varphi\end{pmatrix}$, where $\varphi$ is the azimuthal angle, $r$ is the distance to the source point, and $k_{spp}$ is the SPP propagation constant (Fig. 1A and B). After propagating over an azimuthally-varied distance of $r(\varphi) = r_0 + |m|\frac{\lambda_0}{n_{spp}}\frac{\varphi}{2\pi}$ ($0 \ll \varphi < 2N\pi$), the SPP in-plane components interact with the Archimedean spiral grating that subsequently scatters SPPs into collimated single photons propagating away from the surface and carrying a spiral phase profile of $\delta(\varphi) = n_{spp}\frac{2\pi}{\lambda_0}r(\varphi) = m\varphi + \delta_c$, where $r_0$ is the starting radius, $m$ is the arm number with $m < 0$ for counterclockwise (CCW) spirals, $N$ is the number of spiral windings, $\lambda_0$ is the free-space radiation wavelength, $n_{spp} = \frac{\lambda_0}{2\pi}\text{Re}(k_{spp})$ is the effective SPP mode index, and $\delta_c = n_{spp}\frac{2\pi}{\lambda_0}r_0 + \delta_0$ is a constant phase term including the phase shift $\delta_0$ during the SPP scattering. Hence, the polarization state of generated single photons in the far-field can be represented as the OAM superposition state in the circularly polarized (CP) basis (*20*)

$$|\psi_m\rangle = \frac{1}{\sqrt{2}}(|R\rangle|\ell_R = m+1\rangle - |L\rangle|\ell_L = m-1\rangle) \quad (1)$$

showing unambiguously that the right circularly polarized [RCP, $|R\rangle$] and left circularly polarized [LCP, $|L\rangle$] OAM beams with topological charges of $\ell_R = m + 1$ and $\ell_L = m - 1$, respectively, are generated simultaneously, forming a pair of entangled states composed of different SAM and OAM with the total angular momentum being conserved (*20*).

If the CCW spiral grating has only one arm (i.e., $m = -1$), the accumulated phase variation $\delta$ is continuously changed within the full $2\pi$ range with respect to the azimuthal angle $\varphi$ rotating from 0 to $2\pi$ for scattered electric fields that originate from the spiral head to its tail (Fig. 1C). The corresponding far-field intensity, which can be found by Fourier transforming the scattered near field, consists of the LCP (with $\ell_L = -2$) and RCP (with $\ell_R = 0$) components featuring, correspondingly, a doughnut shape (intensity vanishes at the center due to the phase singularity) and Gaussian profile (Fig. 1A). Contrary to the SPPs outcoupling with a trivial bullseye (concentric circular) grating producing the far-field intensity pattern with a 100% spatial overlap between the RCP and LCP components (Fig. 1D), the introduced spiral gratings result in spatial separation between the RCP and LCP components, with the intensity overlap of only 23% in the considered one-arm spiral grating (Fig. 1E). The spatial separation of the entangled RCP and LCP radiation channels with the same total angular momentum bears important implications to high-dimensional quantum information processing (*16*). The simulated Stokes parameters further validate the angle-dependent polarization of the outcoupled photons characterized by two entangled, RCP and LCP, states (Fig. 1F), consistent with the cross-section profiles in Fig. 1E.

According to the theoretical analysis, we designed three typical CCW Archimedean spiral gratings with $m = -1, -3$, and $-5$ to implement our practical OAM photon sources, ensuring well-defined far-field radiation patterns and considerably high quantum efficiencies (*20*). The optimized geometrical parameters of the $m$-armed concentric spiral are starting radius $r_0 = 450$ nm, free-space radiation wavelength $\lambda_0 = 670$ nm, effective SPP mode index $n_{spp} = 1.218$, and number of windings $N = 9$, which is large enough to transform SPPs into outgoing photons. In the case of $|m| > 1$, the adjacent spiral arms are offset with an angle of $\Delta\varphi = 2\pi/|m|$. The concentric HSQ ridges are placed atop an Ag substrate coated with a 20-nm-thin $SiO_2$ spacer that ensures environmental protection of the Ag substrate and effective coupling of ND-NV radiation into SPPs. Figure 2A displays the schematic of the spiral gratings and their corresponding analytical phase distributions, indicating $m$-fold $2\pi$ phase variation along the azimuthal direction. The performance of the proposed OAM photon sources was first investigated with simulations where the QE was modeled as a $z$-oriented electric dipole and positioned 30 nm away from the $SiO_2$ spacer (*20*). After interaction with Archimedean spiral gratings, the radial SPPs excited from QE radiation are converted into well-collimated outgoing photons carrying different SAM and OAM, which are concentrated around the center in the far-field (Fig. 2B), superior to the previous demonstration (*17*). As a result, the simulated collection efficiency through an objective with a numerical aperture (NA) of 0.9 exceeds 97%, even for the most divergent LCP OAM light with $\ell_L = -6$ achieved with the five-armed spiral. With arm number varied from $-1$ to $-5$, the LCP beam maintains a directional doughnut shape with the divergence angle increased, while the RCP beam changes from Gaussian-shaped to doughnut-shaped, whose topological charges can be directly recognized from the simulated phase windings in the CP basis (Fig. 2C). Since the RCP and LCP components share almost 50/50 power of the outcoupled QE radiation (*20*), the OAM beams with higher topological charges look dimmer. Following simulations, the OAM photon sources were fabricated with NDs containing multiple NV centers by using a combined process including thin-film deposition, spin-coating, and electron-beam lithography (*20*). Figure 2D-F shows the top-view scanning electron microscope (SEM) images of the HSQ spiral gratings with ND-NVs precisely positioned. Compared to semiconductor microring lasers (*21*) or metasurface-assisted surface-emitting lasers (*22*) and SAM single-photon sources (*23, 24*), we directly pattern HSQ resists to make spiral gratings with a smaller vertical feature size of 180 nm, thereby avoiding the etching and multiple steps of alignment that may change or damage the emission property of the selected ND-NVs.

To attain the designed OAM emission, we optically pumped the selected ND-NVs with a tightly focused radially polarized beam at a wavelength of 532 nm to produce a strong longitudinal electric field component $E_z$ at the sample surface, which can selectively excite the ND-NV dipole component along the *z*-axis, mimicking the simulation condition (*25, 26*). In Fourier plane, a linear polarizer (LP) and a quarter-waveplate (QWP) were utilized to decompose the far-field radiation into the CP basis, where the fast axis of the QWP is set to be ± 45° while the LP is vertically placed. When the QWP is −45°-oriented, the projected LCP components all have doughnut shapes spatially expanding with the increased arm number (Fig. 3A). Specifically, the measured divergence angles are $\theta_{div}$ = 4.37°, 7.93°, 10.50° for the spirals with arm number of $m$ = −1, −3, and −5, respectively, consistent with analytical calculations. However, the LCP OAM light with $\ell_L = -2$ originated from the one-armed spiral indicates the lowest contrast between the dark center and bright side ring. We attribute this deviation to the imperfect spherical shape of the selected ND, which makes the diamond-air interface inhomogeneous and the subsequently excited SPPs unevenly distributed. If the QWP is rotated to 45°, a Gaussian-shaped ($\ell_R = 0$ with the main lobe at $\theta = 0°$) and other two doughnut-shaped ($\ell_R = -2$ with $\theta_{div}$ = 4.54° and $\ell_R = -4$ with $\theta_{div}$ = 8.64°, respectively) RCP beams are observed with the increase arm number (Fig. 3B). By integrating the far-field intensities in the CP basis, the intensity fractions between the RCP and LCP components are 48/52, 49/51, and 51/49, respectively, approaching the theoretically calculated and simulated values (*20*).

By employing specific NDs with a small fraction containing single NV centers, collimated OAM single-photons could be generated at room temperature. Compared to NDs with multiple NV centers, these single-NV NDs have smaller diameters and lower scattering cross-sections, which can only be deterministically located with fluorescence images (*20*). After locating several fluorescent ND-NV candidates, a Hanbury Brown and Twiss interferometer was used to determine the ND-NV with single QE characteristics. Before fabrication (the uncoupled bare single ND-NV), the fitted second-order intensity correction function $g^2(0)$ of the selected single-vacancy ND is only ~ 0.16, indicating high single-photon purity (*20*). As anticipated, the far-field intensities projected to CP basis are homogeneously distributed with near-identical patterns in Fourier plane, resulting from the radial SPPs coupled from the QE with the dipole moment normal to the surface. After fabrication of the one-armed HSQ spiral grating around the pre-characterized ND containing a single NV center (the coupled configuration), the auto-correlation $g^2(0)$ is slightly increased to ~ 0.22, maintaining the intrinsic single-photon feature at room temperature (*20*). Impressively, the decomposed LCP beam carries OAM with a topological charge of $\ell_L = -2$ and features a doughnut shape while the RCP component possesses pure SAM and exhibits a Gaussian distribution (Fig. 4A and B). The measured mode purities of the decomposed LCP and RCP components are found to be 61.99% and 66.56%, respectively (*20*). Compared to the simulated values, the mode purity degradation is mainly associated with the broadband spontaneous emission of the ND-NV, the dispersion of the spatial light modulator, the internal off-center location of the NV in the ND, and fabrication imperfections. The HSQ spiral grating also provides a feedback environment for the ND-NV, which could enhance the emission and shorten the lifetime to some extent (*20*). Despite moderate mode purities, this ultrathin planar device effectively reduces integration complexity and emits well-collimated high-purity single photons carrying SAM and OAM superposition states at room temperature.

The generated single photons carry SAM and OAM superposition states and could form two spatially-separated entangled radiation channels with distinctly different polarization properties without any external quantum sources (*15, 27*). We first characterized the correlation between two separated entangled channels on a single photon. As shown in Fig. 4D, a polarizing beam splitter (PBS) was used to split single photons into two separate channels with specific polarization states. The LCP channel and RCP channel carrying OAM and SAM are antibunching with $g^2(0) \approx 0.15$ (Fig. 4C). To show the entanglement between SAM and OAM and validate the Bell state of the generated single photons, we performed full quantum state tomography (QST) to recovery the density matrix by conducting 16 projection measurements with different polarization and OAM bases (*15, 20*). The simulated and measured density matrices of the Bell state $|\psi\rangle = \frac{1}{\sqrt{2}}(|R\rangle|\ell_R = 0\rangle - |L\rangle|\ell_L = -2\rangle)$ are well recovered (Fig. 4E and F), with fidelities approaching 0.991 and 0.656, respectively, indicating the entanglement between SAM and OAM states. Noticeably, the decreased fidelity in the experiment is directly linked to the measured mode purities, primarily limited by the broadband emission spectrum of NV centers, which can be efficiently improved with group IV color centers in NDs possessing sharp zero-phono lines, such as GeV centers(*28*). Importantly, the emitted single photons carrying entangled SAM and OAM states are highly stable in tracing duration up to 20 mins without any operation (*20*).

Concluding, we have demonstrated a room-temperature on-chip integrated OAM single-photon source that emits well-collimated and high-purity single photons carrying entangled SAM and OAM states. The OAM single-photon source is realized by efficiently extracting in-plane diverging SPPs excited from a single ND-NV *via* the HSQ Archimedean spiral grating that impacts spatial phases, which exhibits advantages of surface-confined configuration, reduced fabrication process, and higher collection efficiencies. Our design makes spatially-separated entangled radiation channels addressable by directly streaming single photons with entangled SAM and OAM states without any external sources, paving the way for advanced quantum photonic superdense information processing on a chip (*29*). This compact platform is also feasible for further development of ultrafast and indistinguishable single photons (*30, 31*) with controlled OAM states at room temperature and may open new fascinating perspectives for high-dimensional, large-scale, and integrated photonic quantum systems (*32-35*).

**Acknowledgments:**

C.W. and F.D. appreciate the helpful discussion with Prof. Jianwei Wang from Peking University.

**Funding:** The work was funded by the Villum Fonden (award in Technical and Natural Sciences 2019, grant nos. 00022988, 37372, and 35950). Z.M.W and C.W. acknowledge the support from the National Key Research and Development Program of China (2019YFB2203400), the "111 Project" (B20030), the UESTC Shared Research Facilities of Electromagnetic Wave and Matter Interaction (Y0301901290100201) and the China Scholarship Council (Grant No. 2020023TO014).

**Author contributions:** C.W. and F.D. conceived the concept. S.I.B. and F.D. conceived the theoretical model and C.W. performed numerical simulations. C.W. and Y.H.K. fabricated the samples. C.W., S.K. and D.K. built optical setups and characterized the devices. C.W., S.K., Z.M.W., S.I.B. and F.D. analyzed the data. C.W. and F.D. drafted the manuscript. All authors contributed to editing and preparing the manuscript. F.D. and S.I.B. supervised this project.

**Competing interests:** Authors declare that they have no competing interests.

**Data and materials availability:** All data are available in the main text or the supplementary materials.

**Supplementary Materials (available in the online version of the paper):** Theoretical Modelling, Numerical simulations, Device preparation, Optical and QST measurements.

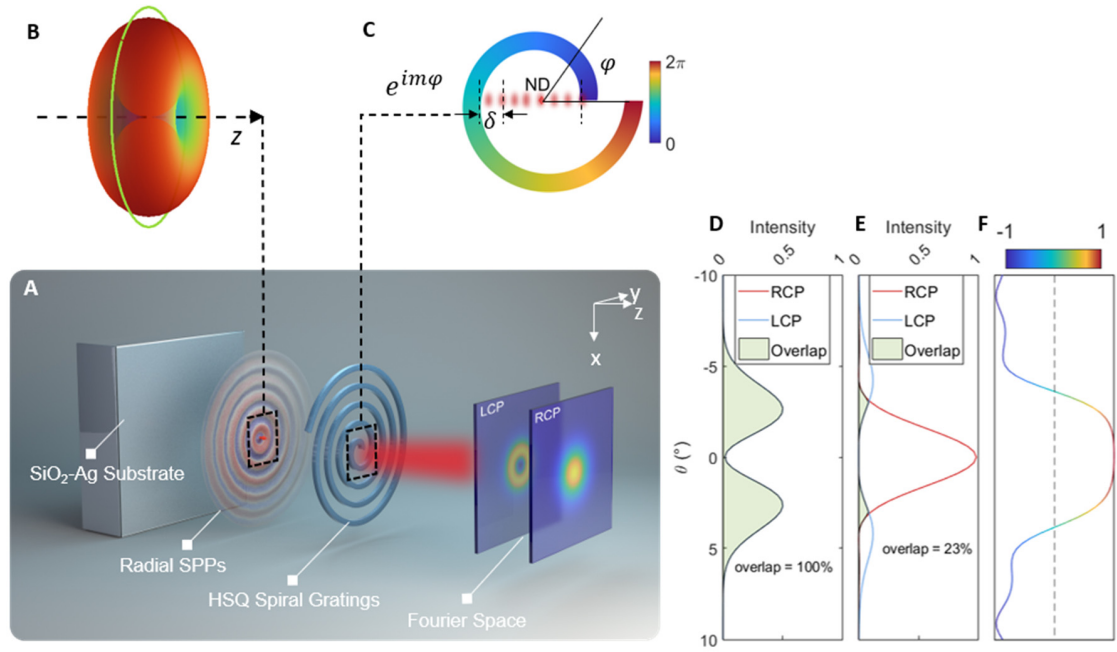

**Fig. 1. Design of the collimated OAM single-photon source.** (**A**) Schematic of the OAM single-photon source operation in layer-by-layer representation: a SiO$_2$-coated Ag substrate, supporting radial SPPs excited by a QE *z*-oriented dipole; an HSQ spiral grating, outcoupling SPPs into a well-collimated photon stream; and decomposed far-field RCP and LCP intensity profiles, from left to right, respectively. (**B**) 3D rendering of the electric field radiation of a *z*-oriented dipole. (**C**) Analytical azimuthal phase distribution generated due to the propagation and scattering of radial SPPs on the first winding of the HSQ spiral. (**D** and **E**) Analytical cross-section profiles of far-field RCP and LCP components produced with radial SPPs being outcoupled by a bullseye grating (D) and one-arm CCW spiral grating (E). The intensity overlap between the RCP and LCP components is 100% and 23%, respectively, demonstrating spatial separation of entangled radiation channels with distinctly different polarization properties. (**F**) Simulated Stokes parameter $S_3$ of RCP (+1) and LCP (−1) components in the total field.

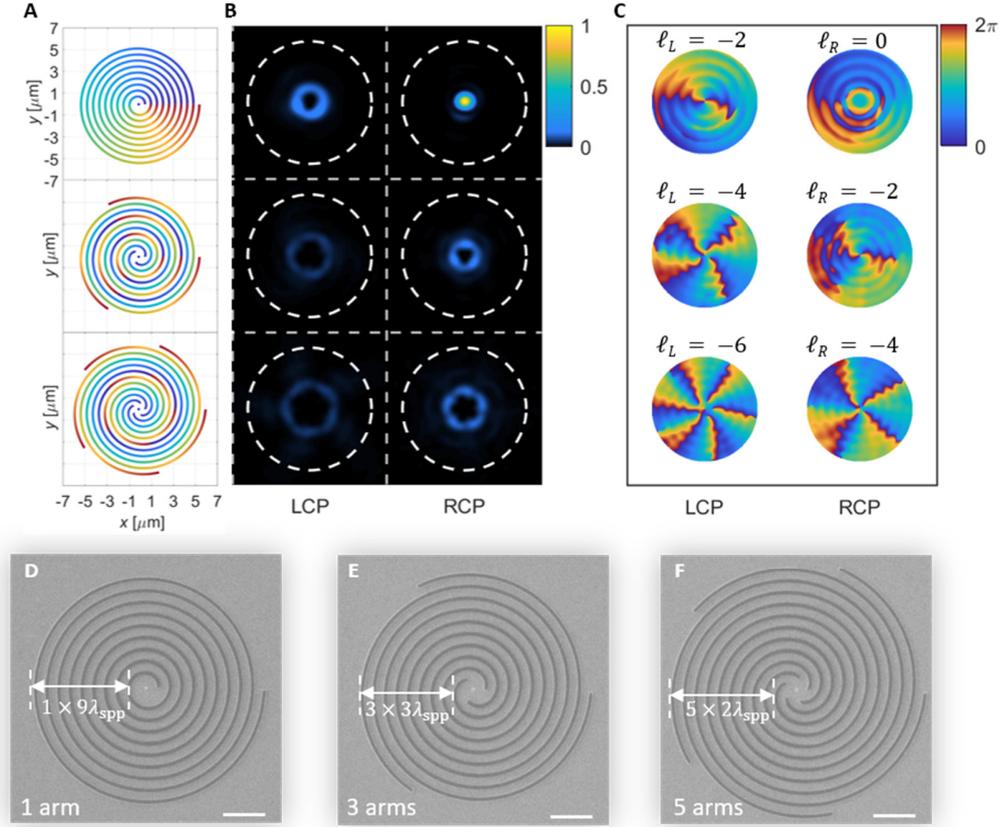

**Fig. 2. Simulation of the collimated OAM photon sources.** (**A**) Schematic of the spiral gratings and their corresponding analytical phase distributions with arm number of *m* = −1, −3, and −5 from top to bottom, respectively. (**B**) Decomposed far-field LCP and RCP intensity profiles normalized to the one-armed spiral grating. The white dashed circles denote the collection angle with NA = 0.2. (**C**) Simulated phase windings in the far-field for corresponding configurations. (**D-F**) SEM images of the fabricated OAM photon sources with arm number of *m* = −1, −3, and −5. $\lambda_{spp} = \frac{\lambda_0}{n_{spp}}$ is the SPP wavelength. The scale bar is 2 μm.

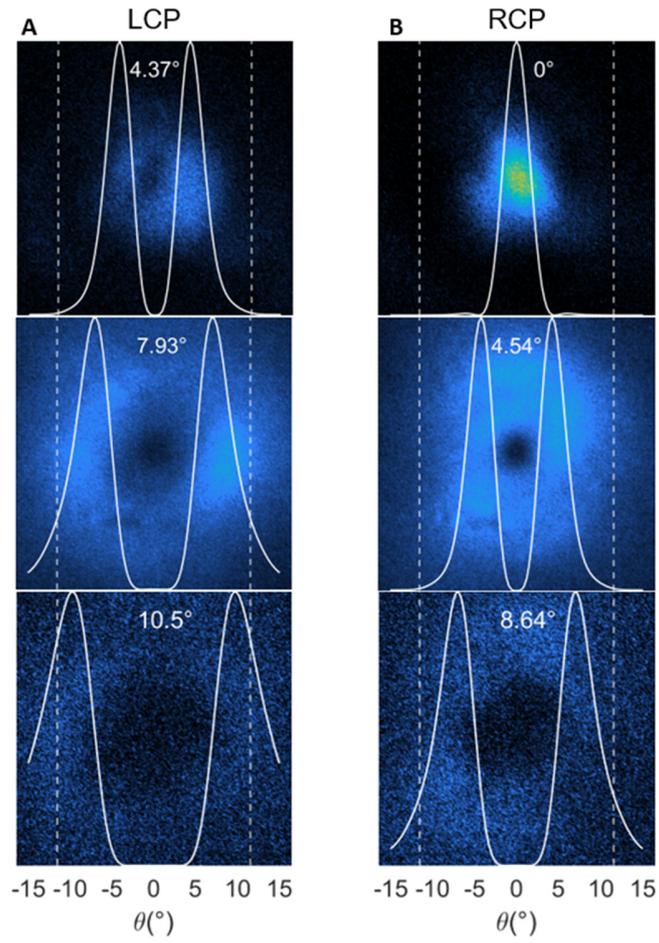

**Fig. 3. Experimental demonstration of the collimated OAM photon sources.** (**A** and **B**) Measured far-field LCP (A) and RCP (B) intensity patterns. Superimposed white curves are the analytically calculated cross-section profiles along the *x*-axis while dashed lines indicate the collection angle with NA = 0.2.

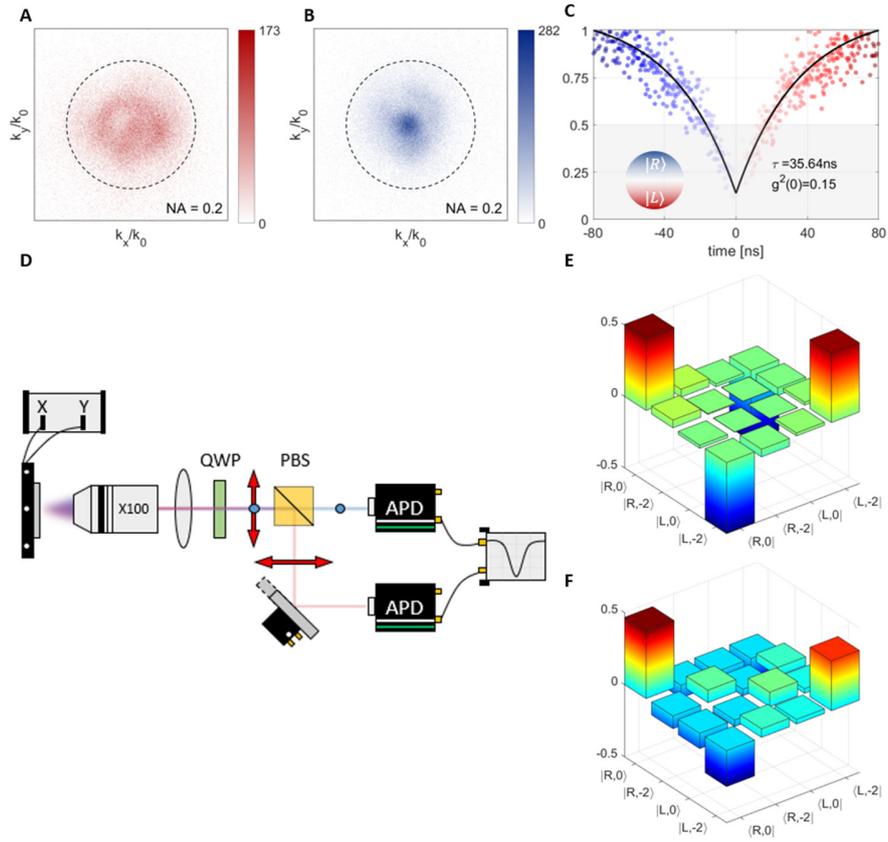

**Fig. 4. Experimental demonstration of the collimated OAM single-photon source.** (**A** and **B**) Measured far-field LCP (A) and RCP (B) intensity patterns with the single-photon source. (**C**) Measured correlation between LCP and RCP channels. (**D**) Simplified representation of a modified Hanbury Brown and Twiss setup for correlation measurement between RCP and LCP channels. (**E** and **F**) Simulated (E) and experimental (F) density matrices for the Bell state |ψ⟩ on single photons.